# Towards Structure-Property-Function Relationships for Eumelanin

*Paul Meredith[1]\*, Ben J. Powell[2], Jennifer Riesz[1], Stephen Nighswander-Rempel[1], Mark R. Pederson[3], and Evan Moore[4]*

\*[1]Soft Condensed Matter Physics Group, University of Queensland Department of Physics, St. Lucia Campus, Brisbane, QLD 4072, Australia
[2]Theory of Condensed Matter Physics Group, University of Queensland Department of Physics St. Lucia Campus, Brisbane, QLD 4072, Australia
[3]Center for Computational Materials Science, Naval Research Laboratory, Washington, D.C. 20375, USA
[4]Lawrence Berkeley National Laboratory, Berkeley, CA 94720, USA

\*meredith@physics.uq.edu.au

We discuss recent progress towards the establishment of important structure-property-function relationships in eumelanins – key functional bio-macromolecular systems responsible for photo-protection and immune response in humans, and implicated in the development of melanoma skin cancer. We focus on the link between eumelanin's secondary structure and optical properties such as broad band UV-visible absorption and strong non-radiative relaxation; both key features of the photo-protective function. We emphasise the insights gained through a holistic approach combining optical spectroscopy with first principles quantum chemical calculations, and advance the hypothesis that the robust functionality characteristic of eumelanin is related to extreme chemical and structural disorder at the secondary level. This inherent disorder is a low cost natural resource, and it is interesting to speculate as to whether it may play a role in other functional bio-macromolecular systems.

# 1. Introduction

The melanins are a broad class of functional bio-macromolecule found throughout the biosphere.[1] Due to their strong absorption of visible light they have long been thought to act as pigments, serving an important photoprotective role in the skin, eyes and hair of humans and many other animals. However, their presence in the *substantia nigra* of the brain stem and other areas not exposed to sunlight suggests other functions, and it has been recently proposed that melanins play an even stronger role in immune response than in photoprotection.[2] Paradoxically, precursors of eumelanin and pheomelanin (the two predominant forms in humans) are implicated in the development of melanoma skin cancer.[3] Additionally, deficiencies in neuromelanin (a hybrid of eumelanin and pheomelanin) have been found in the brain stems of patients suffering various neuro-degenerative diseases such as Parkinson's.[4] Clearly, these bio-macromolecules are important from a biological and medical perspective, a fact that has motivated many decades of research from biochemists, pigment cell biologists, molecular biologists and the broader biology community.

More recently, these systems have attracted the attention of molecular biophysicists and the soft matter and functional organic materials communities.[5] Eumelanin in particular, possesses an intriguing and rather unique set of physicochemical properties: strong broad band UV and visible absorption; non-radiative conversion of photo-excited electronic states approaching unity (extremely low radiative quantum yield)[6]; powerful anti-oxidant and free radical scavenging ability[7]; and probably most bizarrely of all, electrical conductivity and photoconductivity in the condensed phase[8,9]. In relation to the last two properties, it has even been speculated that these systems may be bio-organic semiconductors.

Melanins have proven to be an intractable system to study, in part due to their strong binding with a protein host under *in vivo* conditions. It is not understood whether this protein host is a left-over from the bio-synthesis, or whether it is crucial for the functionality. Strictly speaking, the term "melanin" should encompass both the chromophore and the associated protein. However, it has become the norm to use the term to refer only to the chromophore (the nominally functional part). The vast majority of studies on the natural system remove (or ignore) the protein, and all studies on synthetic melanins involve the chromophore only. Unfortunately, the chromophores themselves are virtually insoluble in most common solvents, and have defied systematic attempts at characterisation by standard analytical methods. The predominant strategy in the field is to study the photochemical and photophysical properties of isolated natural models or synthetic analogues without recourse to any consistent structural model. These difficulties have resulted in a number of key questions remaining unanswered, including:
1. What are the appropriate primary and secondary structural models for each melanin type? (Traditionally, these systems have been viewed as extended heteropolymers[10], but it has been suggested recently that aggregates of smaller oligomeric structures may be a more accurate model[11].)
2. How does supramolecular organisation impact / control emergent macroscopic properties such as optical absorption, electrical conductivity, etc.?

3. How do these structures compare in synthetic and natural systems, and are changes at the primary and secondary level responsible for altered functionality in disease states?

In addition to this list of difficult and intriguing questions, one may also speculate as to whether fundamental knowledge of key structure-property relationships will allow us to use molecular engineering approaches to create new bio-inspired functional soft materials based upon melanins.

In this article we describe recent advances in answering some of these important questions for the case of eumelanin. These systems are known to be macromolecules of 5,6-dihydroxyindole (DHI, also known as HQ) and 5,6-dihydroxyindole-2-carboxylic acid (DHICA) and their various oxidised forms (Fig. 1).[1] In particular, we focus on the powerful combination of detailed molecular spectroscopy and quantum chemical calculations of electronic and vibrational structure. We will argue that there is mounting evidence that extreme chemical heterogeneity (in the secondary and perhaps primary structure as well) plays a key role in defining the macroscopic properties of these systems, and indeed, may be the reason for the characteristic robust functionality in natural melanins.

## 2. Optical Spectroscopy: Absorption, Emission and Excitation

*a) Absorption*: All melanins are known to have characteristic broad band absorption in the UV and visible[12,6]. Eumelanin (from both natural and synthetic sources) has a monotonic absorbance profile (Fig. 2). Note that in contrast to most organic systems, there are no distinct chromophoric bands, the profile fits a simple exponential, the absorbance rises steeply in the ultra-violet (a very useful property for a human photo-protectant), and no low energy limit has yet been discovered. The prevailing paradigm in the field is that melanins are indeed disordered organic semiconductors.[13,14] This model naturally explains the observed broad band absorbance, electrical conductivity and photoconductivity[15]. The lack of an apparent gap in either solution absorbance measurements or solid state transmittance and reflectance measurements is somewhat inconsistent with this picture, but might be explained by invoking scattering and / or the band edge tails characteristic of amorphous systems. Suffice it to say that absorption is not a sensitive probe of underlying structure in these systems, and as we shall see later, a much less sophisticated model can be used to explain its broad, smooth form.

*b) Emission*: There are conflicting reports concerning the photoluminescence (fluorescence) of melanins[6,16,17]. It is not uncommon to read in the literature that "melanins do not fluoresce". This is not strictly true; eumelanin at least does emit radiatively when stimulated by UV and visible light, but its photoluminescence quantum yield is tiny[18,19,20]. Radiative quantum yields have been measured by applying a procedure to account for probe beam attenuation and emission re-absorption.[19] Using this procedure, it is also possible to reconstruct the form of the emission; these results have led to several unexpected observations. Firstly, the form of the emission was excitation energy dependent – in direct contradiction of Kasha's rule[21]. Fig. 3 shows how the emission full width at half maximum is reduced and the peak position red shifted as excitation energy decreases (wavelength increases). Secondly, this system violates the mirror image rule for organic chromophores: namely that the form of the emission spectrum should be the approximate mirror image of the absorbance (excitation). Thirdly, there appears to be a constant low energy limit to the emission (~2.0 eV, 620 nm), which could indeed be associated with some minimum energy transition gap in the system. Radiative quantum yields were determined by normalising corrected emission spectra against the absorptivity at the excitation wavelength.[19] These yields were also shown to be dependent on excitation wavelength (Fig. 4).

These studies have recently been extended to create a full "radiative quantum yield map"[19], showing the fate of each absorbed photon with respect to any emission (Fig. 5). This map confirms the complex, excitation energy dependence of the emission, and the presence of a high wavelength (low energy) bound on the emission. It appears that this bound corresponds to some form of minimum energy gap in the system. However, it is clear that the radiative decay pathways in this system are far more complex than one could expect for a single organic chromophore. In this map, a high energy limit to the emission was also observed. In contrast to the low energy limit, this is not due to a lack of transitions with corresponding energy differences because melanin absorbs strongly at these energies. Instead, it seems that upon excitation at high energies, eumelanin preferentially de-excites (non-radiatively) to lower orbitals and fluoresces within a fixed energy range. The extremely low radiative quantum yield values are consistent with pump probe and time resolved studies which show very efficient non-radiative coupling[20,22].

*c) Excitation*: The available literature on the excitation characteristics of eumelanin is even more sparse and inconsistent than the emission characteristics.[18,23,24] The same procedure to correct for probe beam attenuation and emission re-absorption can be used to re-construct the photoluminescence excitation profiles[18]. This has recently been done for synthetic eumelanin (a series of corrected spectra are shown in Fig. 6). The excitation spectra (although having the same general form as the absorption spectra) contain a wavelength invariant feature at ~365 nm (3.4 eV) irrespective of the emission energy. This phenomenon is also apparent in the radiative quantum yield map (Fig. 5). In the context of a simple, single organic chromophore, such a feature would indicate that a particular transition pumped by 365 nm (3.4 eV) radiation decays via a pathway with a higher radiative probability than transitions from surrounding states, since there is no corresponding peak in the absorbance spectrum. This in turn implies that the particular electronic excited state in question is more weakly coupled to vibrational modes of the system, and it was suggested that this feature is due to DHICA monomers at the extremities of the melanin compound.

We can summarise the key findings from these optical spectroscopy experiments as follows:
1. The absorption in eumelanin is broad and monotonic extending from the UV to the near-IR.
2. The emission in these systems is highly atypical of a single organic chromophore. In addition to violating Kasha's rule and the mirror image rule, it shows a low energy (high wavelength) bound at ~2 eV, which may correspond to a minimum gap in the system and a high energy bound which may represent preferred de-excitation pathways.
3. The radiative quantum yield is extremely small (order of $10^{-4}$).

4. The excitation spectra contain a notable, fixed (with respect to emission) peak at 365 nm that is not present in the absorbance spectrum.
5. The emission-excitation landscape as revealed by the specific quantum yield map is very complex, and is again atypical for a single organic chromophore.

Given the above findings, we have been forced to consider whether eumelanin may in fact consist not of a single identifiable chromophore but rather a heterogeneous ensemble of many chemically distinct chromophores each with a different HOMO-LUMO gap (the gap between the highest occupied molecular orbital and the lowest unoccupied molecular orbital). In the next section, we describe some recent quantum chemical results which seem to support this "chemical disorder" hypothesis.

## 3. Quantum Chemistry: First Principles Density Functional Calculations of the Building Blocks of Eumelanin

The most relevant calculation for comparison with the optical studies described above is absorption. Thus the HOMO-LUMO gap provides a simple, single parameter to compare with experimental results and is often the primary object of computational studies. Unfortunately, most computationally inexpensive methods can only provide semi-quantitative values for the HOMO-LUMO gap (i.e. trends and orders of magnitude are reliable, but exact values may be incorrect by 20% or more). For example, density functional theory (DFT)[25-29], which has been the mostly widely used method for theoretical studies of melanin precursors, suffers from the notorious band gap problem[26]. Nevertheless, both the difference of self consistent fields ($\Delta$SCF) and time-dependent DFT (TDDFT) methods have provided reasonable estimates of the HOMO-LUMO gap; a summary of such calculations for the molecules shown in Fig. 1 is given in Table 1. An interesting and rather unexpected fact emerges: simple de-protonation of the 5 and / or 6 positions (transition from the phenolic to ketone forms) produces a dramatic red shift (reduction in energy) of the gap. Stated simply, the oxidised IQ and SQ monomeric forms are predicted to be coloured, whilst the protonated form (HQ and DHICA) are predicted to be non-coloured. At first sight, this may be a simple prediction to test – synthetic methods of preparing isolated monomers of DHI are available. However, the molecule is extremely unstable in an oxidising environment, and will spontaneously polymerise under any conditions apart from those in which oxygen and light have been completely excluded. This makes obtaining a reliable absorption spectrum somewhat difficult, but Zhang *et al.*[30] have published a DHI spectrum in acetonitrile from which a HOMO-LUMO gap of ~4.0 eV can be extracted which agrees with the computed value (3.6 eV).

The relative energies of HQ, IQ and SQ calculated in reference 26 predict that IQ and SQ are very unstable *in vacuo*. This in fact appears to be the case even in a solvent, and no experimental gap energies have been measured for these species. De-protonation of the 5 and / or 6 positions of DHICA has a similar red shifting effect[27], and, *in vacuo*, the oxidised forms are energetically even less favourable in this case. Again, the agreement between the predictions of the DHICA gap energy[27] and experimental values[31] (in aqueous solution) is good (within 20%) especially considering the band gap problem and the fact that calculation pertains to molecules *in vacuo*. The electron densities shown in Fig. 7 give some indication as to why the gap shrinks in the oxidised forms. The HOMO and LUMO electron densities in IQ are less delocalised (because of the electron withdrawing effect of oxygens in the 5 and / or 6 positions, and the loss of aromaticity) relative to HQ. This decreased delocalisation would indeed be expected to reduce the gap energy.

These calculations have been extended to dimers and larger oligomers[29]. Our own calculations[32], and those of Stark *et al.*[29] and Il'ichev and Simon[25] predict that further red shifting of the HOMO-LUMO gap can occur upon oligomerisation. Stark *et al.*[29] have also predicted that "stacking" of oligomeric sheets can also cause modification of the gap. However, the problem of the secondary structure of eumelanin rears its head here. Without knowledge of how oligimerisation occurs, these calculations are speculative. The number of plausible oligomers (for example ~50 different plausible dimer structures can be made by combining HQ, IQ and SQ) renders a systematic study infeasible partly because of the computational cost, but mainly because of the difficulty of drawing relevant connections between theory and experiment.

Despite this last caveat, the capacity to perform accurate first principles calculations on eumelanin building blocks and likely reaction products is a valuable tool in attempts to understand the secondary structure and interpret spectroscopic findings. The findings discussed above can be summarised as follows:
1. Oxidation of both DHI and DHICA to IQ and SQ forms causes dramatic red shifting of the HOMO-LUMO gap.
2. Oligomerisation and stacking can cause further red-shifting of the gap energies.
3. A large number of macromolecules with different proportions of oxidised and reduced sub-units (and with different bonding configurations) can form and appear energetically stable. Given points 1 and 2, this leads to the conclusion that a range of chemically distinct species can form, and that these species will possess a range of HOMO-LUMO gap energies.

The proposition outlined in point 3 suggests that an oligomeric rather than polymeric secondary structure is appropriate. Irrespective of the size of the "eumelanin supramolecule", and indeed, irrespective of whether stacking plays any significant role, the crucial question is: could an ensemble of similar, but chemically distinct species explain the observed optical properties?

**4. Chemical Disorder vs. Optical Properties?**

Recall that the optical absorption of eumelanin is monotonic and fits a simple exponential function with respect to wavelength (Fig. 2). Ground to excited state transitions in organic chromophores in solution are characterised by inhomogeneously broadened Gaussian line shapes (in both energy and wavelength). A linear superposition of several line shapes (each with different peak position and intensity) would indeed produce a broadened line shape. In the limit, where the individual line shapes are not resolvable, this could produce a monotonic profile. The intensity of a peak in an absorbance spectrum is dependent upon both the concentration of the species producing that peak and the transition dipole moment of the particular transition associated with the peak.

Fig. 8 demonstrates how a smooth, monotonic exponential function can be generated by a relatively small number of broadened Gaussian transitions in wavelength space. In this simple illustration, a linear combination of 11 Gaussians with peak spacings of 50 nm and full width at half maxima (FWHM) of 90 nm (a value typical for an organic chromophore in a polar solvent) produces the required smooth exponential function.

Reduction of the FWHM to approximately the peak spacing creates "ripples" in the profile. Modification of the relative Gaussian peak heights changes the form of the exponential, and indeed, produces multi-exponential behaviour in the high and low wavelength limits. This qualitative illustration shows how a simple linear combination of individual absorption features can lead to a featureless exponential profile. The calculations summarised in Section 3 indicate that eumelanin-like species with HOMO-LUMO gaps in the UV, visible, and near-IR can form and they are stable. A range of bonding configurations and redox states do create a range of gaps. Therefore, an ensemble of similar but chemically distinct species can explain the observed monotonic, broad band absorbance and an amorphous, semiconducting band structure is not required.

Likewise, the complex emission and excitation behaviour exhibited by these systems can be explained using the chemical disorder model. Different excitation energies will selectively pump different species within the ensemble. Additionally, individual components, being chemically distinct, would likely have different radiative quantum yields and be present at a range of concentrations. The low energy limit at 2 eV revealed in both emission and excitation data would correspond to the species with the lowest energy HOMO-LUMO gap. Recent solid state photo pyroelectric measurements[5] seem to confirm that some form of gap does begin to emerge at ~ 1.7 eV (the calculated HOMO-LUMO gap of IQ is 2.0 eV). At the UV end of the excitation spectra and quantum yield maps, the distinct 365 nm feature (3.4 eV) is likely associated with those macromolecules on the periphery of the melanin compound and may reflect the presence of DHICA (the calculated HOMO-LUMO gap for HQ is 3.6 eV).

Lastly, one may speculate that the relatively small radiative quantum yields exhibited by eumelanins could also be a result of their inherent chemical heterogeneity. A range of species with progressively lower HOMO-LUMO gaps would create the conditions for emission-reabsorption cascade, ultimately leading to near unity non-radiative conversion of any absorbed photon. Of course, strong intra and intermolecular coupling (as well as solvent coupling) are also likely sources of efficient fluorescence quenching. These factors are amplified in complex supramolecular systems (c.f. the quenching of the green fluorescent protein chromophore when removed from its native protein cage[33]).

Hence, it would appear that the chemical disorder model can explain all of the optical features that we observe in spectroscopic studies of synthetic eumelanin.

**5. Conclusions and Future Perspectives**

In this paper we have described some key recent results concerning the structure-property relationships of eumelanin. In particular, we have focused on the secondary structural model, and how this relates to optical properties. Significant new insights have been gained by coupling quantum chemical calculations with spectroscopic measurements on model eumelanin systems. These results indicate that significant chemical heterogeneity at the primary and especially secondary level could explain many of the observed properties. Eumelanins appear to be composed not of a single chromophore type (whether they be extended polymeric or smaller oligomeric systems), but rather of ensembles containing a range of chemically distinct macromolecules. We call this the "chemical disorder model".

Melanins (and eumelanin in particular) display remarkably robust structure-property and structure-function relationships. For example, a number of apparently dissimilar (at the primary chemical level) systems possess broad band absorbance, low radiative yield and anti-oxidant behaviour. Most melanins are also very stable. In this sense, we could consider the chemical disorder and heterogeneity at the secondary / supramolecular level as being a low cost resource utilised by nature to create robustness. Small amounts of heterogeneity can be found in other functional bio-macromolecular systems as well. For example coupling heterogeneity in the antenna chlorophylls of PSI of cyanobacteria is thought to red shift and broaden the antenna absorption cross-section[34]. However, the dominant paradigm in molecular biology is that exquisite and precise structuring at the primary, secondary and tertiary levels dictates function. It is therefore interesting to speculate as to whether the degree of *chemical* disorder inherent in eumelanin is a unique feature of these systems, or whether it arises elsewhere in biology.

Because of the inherent heterogeneity in the natural material, and the difficulties associated with chemical analysis of the full macromolecular system, a model "bottom-up" molecular approach may be the way forward for melanins. The ultimate goal of a full mapping of the structure-property relationships in eumelanin therefore seems within the reach of a combined experimental and computational program on synthetic systems. Given what has recently been learned about heterogeneity in these systems, we believe a more appropriate fundamental question to ask may be: "*what is the minimum level of supramolecular organization and chemical heterogeneity required to explain the physio-chemical properties of melanins*".

## Acknowledgements

The work at the University of Queensland was part funded by the Australian Research Council, and by the University of Queensland Strategic Initiative in Condensed Matter. MRP was supported in part by Office of Naval Research and the Department of Defence HPC CHSSI initiative. The authors would like to acknowledge valuable input from Paul Burn (Oxford), Ross McKenzie (UQ), Tad Sarna (Jagiellonian University), Peter Parsons (Queensland Institute for Medical Research), and Tunna Baruah (NRL and Georgetown University).

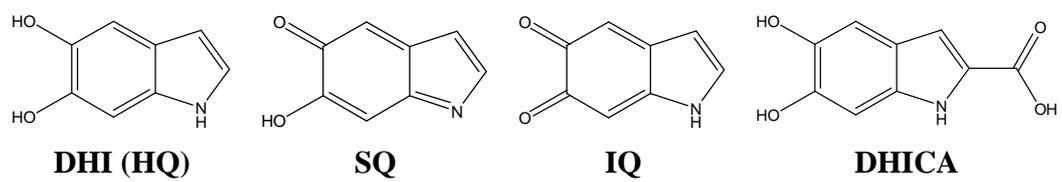

**Fig. 1** 5,6-dihydroxyindole (DHI), its oxidized forms (IQ and SQ), and 5,6-dihydroxyindole, 2-carboxylic acid (DHICA): the key monomeric building blocks of eumelanin.

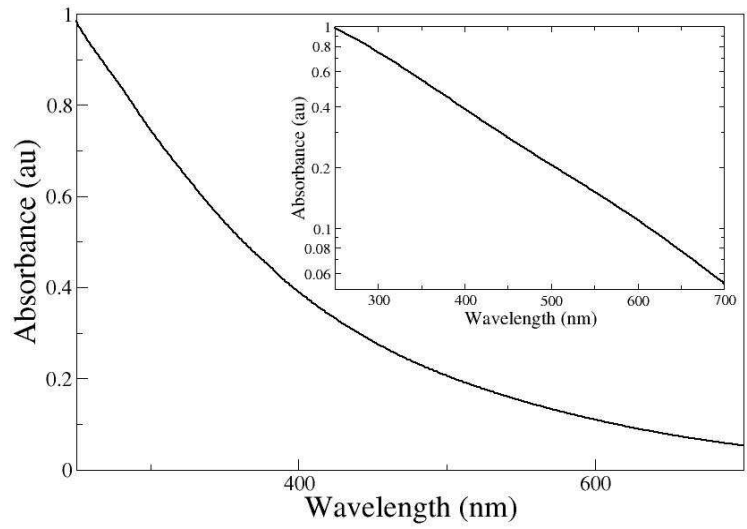

**Fig. 2** Absorbance as a function of wavelength of a typical synthetic eumelanin aqueous solution (0.0025%/wt). The same data is shown on a log-linear plot in the inset demonstrating the excellent fit of the raw data to an exponential form.

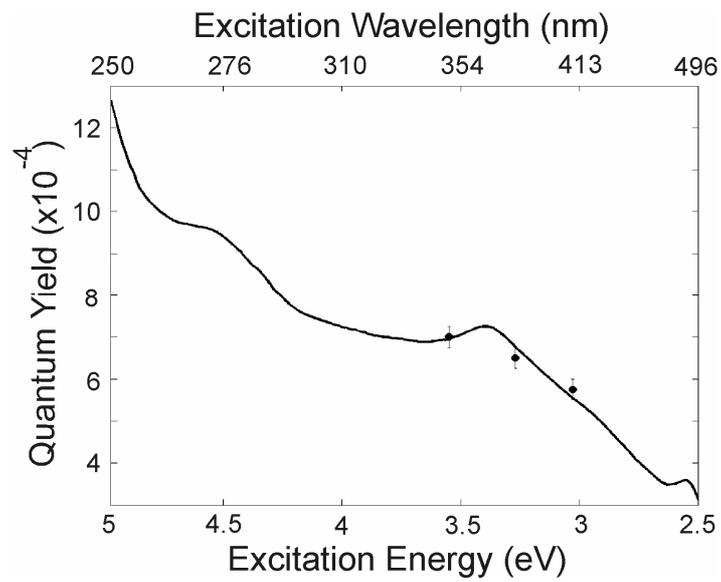

**Fig. 3** Radiative quantum yield as a function of excitation energy and wavelength for synthetic eumelanin – dots represent data from reference 6, and solid line from reference 19.

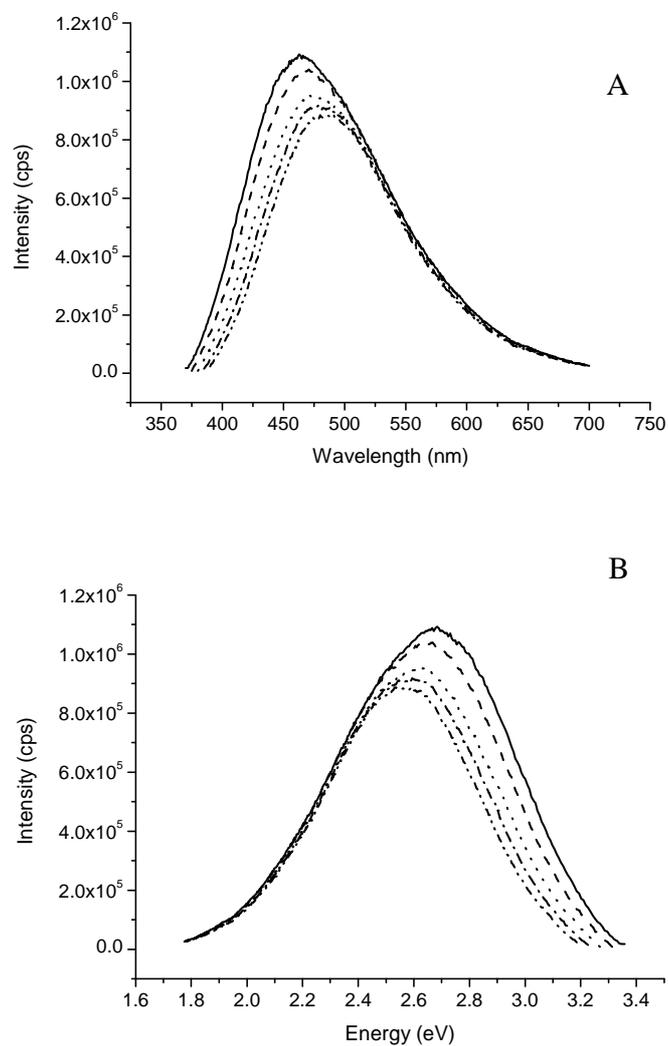

**Fig.4** Fluorescence emission of a synthetic eumelanin solution (0.0025% by weight) a) as a function of wavelength and b) as a function of energy at a number of excitation wavelengths (energies) from 360 nm (solid line) to 380 nm (inner dot-dashed line) in 5 nm intervals. Data from reference 6.

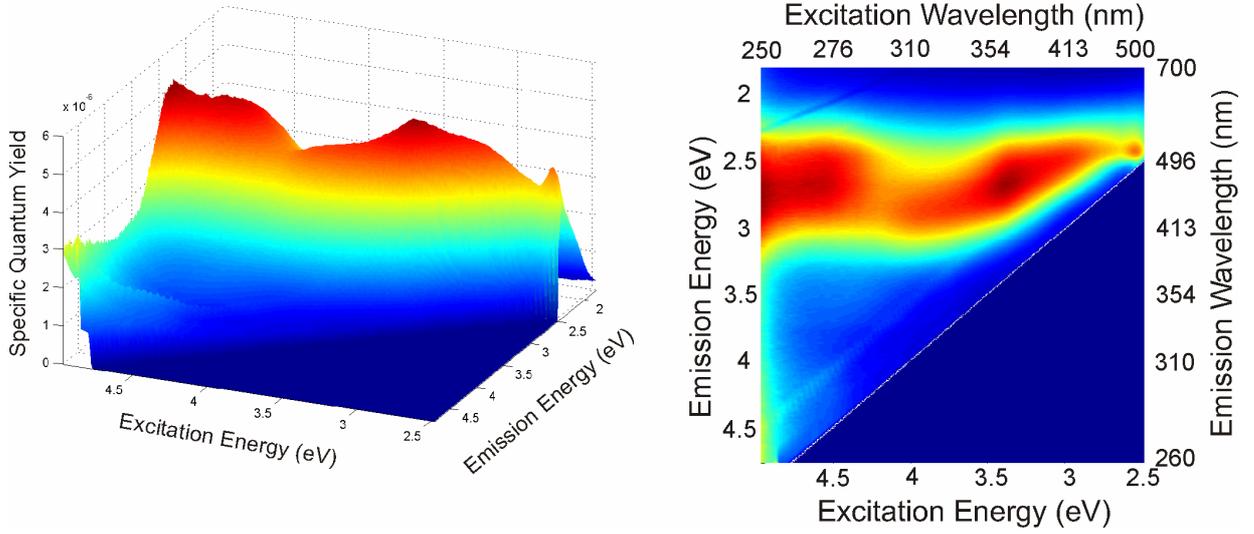

**Fig. 5** Specific quantum yield map for synthetic eumelanin: the fraction of photons absorbed at each excitation wavelength that are emitted at each emission wavelength. Two peaks are evident and limiting values at high- and low-emission are observed. These maps show the fate of each photon absorbed relative to the limited emission that occurs in eumelanins. The specific quantum yield can be calculated from:

$$Q(\lambda_{ex}, \lambda_{em}) = \frac{I_d^*(\lambda_{ex}, \lambda_{em})}{C\left(1 - e^{-\alpha(\lambda_{ex})d_{ex}}\right)} \qquad \text{eq.1}$$

Where $I_d^*$ is the measured emission intensity corrected for probe beam attenuation and re-absorption as per reference 6, $\lambda_{ex}$ and $\lambda_{em}$ are respectively the excitation and emission wavelengths, $\alpha$ is the absorption coefficient as a function of the excitation wavelength, and $d_{ex}$ is the width of the excitation volume. The traditional radiative quantum yield is then given by:

$$\phi(\lambda_{ex}) = \frac{1}{C} \frac{\int I_d^*(\lambda_{ex}, \lambda_{em}) d\lambda_{em}}{1 - e^{-\alpha(\lambda_{ex})d_{ex}}} \qquad \text{eq.2}$$

Note that the factor $1/C$ is a normalising parameter dependent only on the system geometry and the detector sensitivity. This can be found by measurement of a quantum yield standard such as quinine or rhodamine 6G. Data from reference 19.

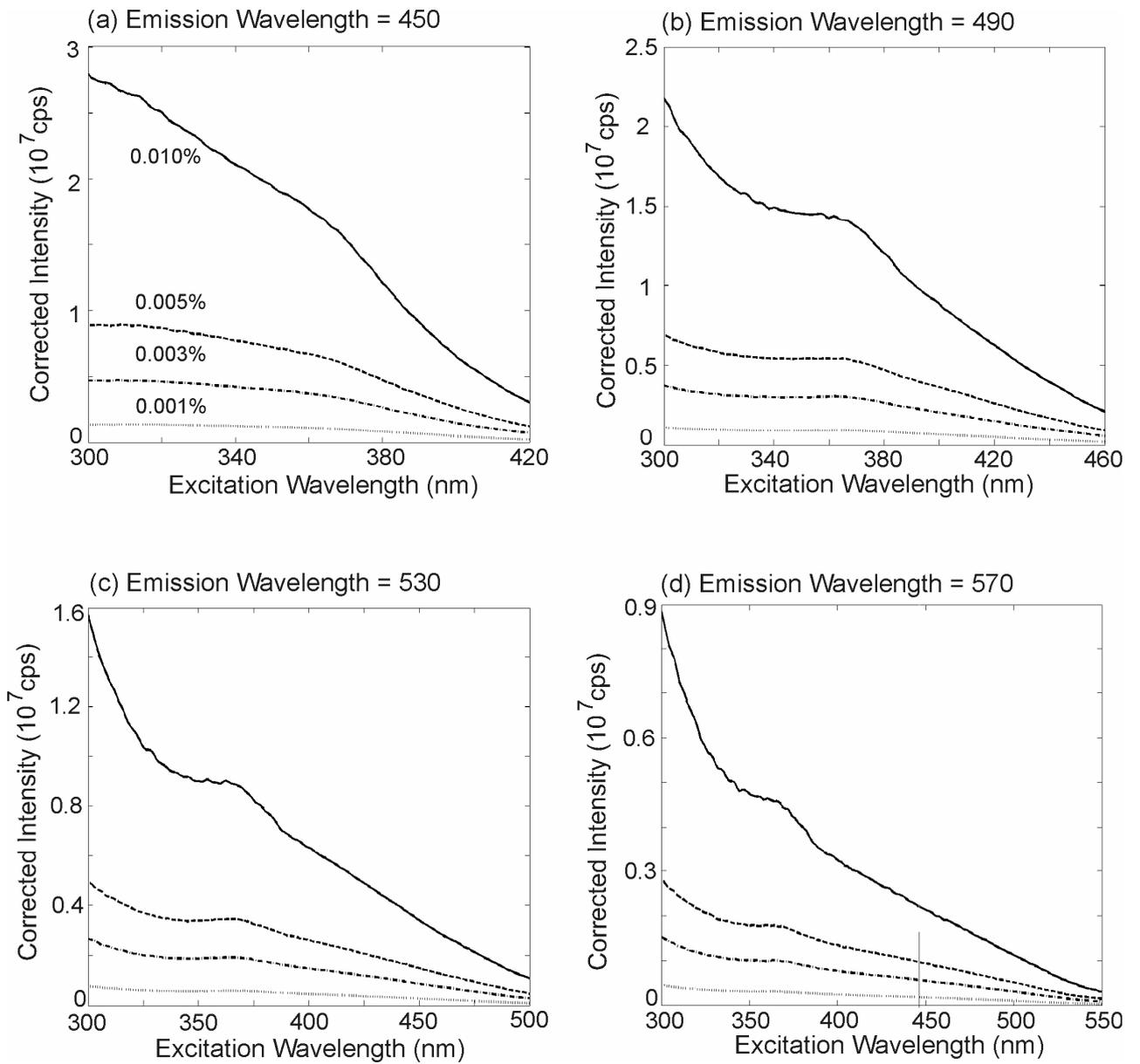

Fig. 6 Corrected excitation scans for four emission wavelengths: a) 450 nm, b) 490 nm, c) 530 nm and d) 570 nm and for four synthetic eumelanin concentrations: 0.001% (dot), 0.003% (dot-dash), 0.005% (dash), and 0.010% (solid) by weight. Data taken from reference 18.

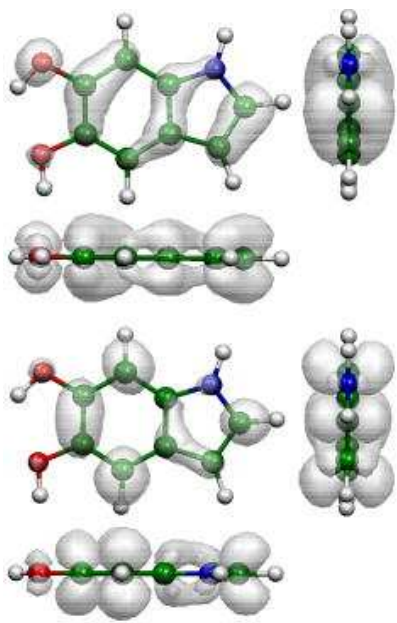 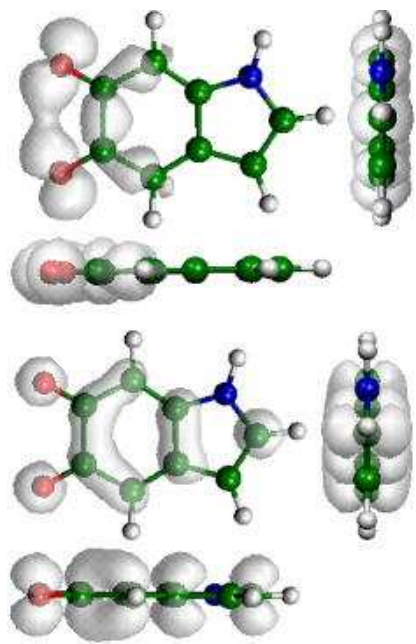

**Fig. 7** The electron densities in the HOMO (top) and LUMO (bottom) states for a) HQ (DHI) and b) IQ. It can be seen that both sets of orbitals have π-character. Data from reference 26.

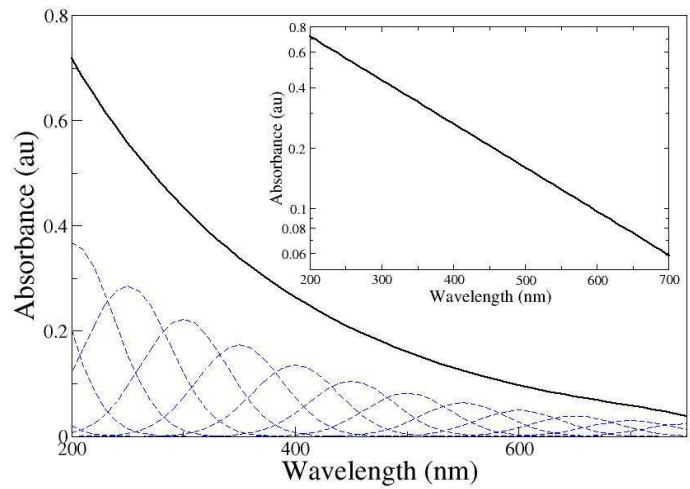

**Fig. 8** The broad band monotonic profile characteristic of eumelanin absorbance. The absorbance spectrum (solid black line) has been simulated using a linear combination of 11 Gaussians, FWHM = 90 nm, peak separation = 50nm (shown as dashed blue lines). The intensity, *I*, of the Gaussians is given by the function $I = \exp(-\lambda/C)$, where $\lambda$ is the peak wavelength, thus the only free parameter is the constant *C*, which we take to be 200 nm. The same curve is shown on a log-linear plot in the inset.

| Method/Functional | IQ | SQ | HQ | DHICA |
|---|---|---|---|---|
| DFT-ΔSCF/PBE | 2.02 | 1.12 | 3.61 | 3.04 |

**Table 1** HOMO-LUMO gaps in eV for DHI (HQ), SQ, IQ and DHICA. Data from references 26 and 27. These density functional theory calculations were performed using the Perdew-Burke-Ernzerhof (PBE) exchange correlation functional and the ΔSCF method (difference of self consistent field).